\documentclass[preprint]{aastex}

\slugcomment{Accepted on February 11, 2013, for the publication of ApJ}

\usepackage{color}

\shorttitle{High-resolution Imaging of the Gegenschein \\and the Geometric
Albedo of Interplanetary Dust}
\shortauthors{Ishiguro et al.}

\makeatletter
\AtBeginDocument{\@ifpackageloaded{natbib}{\ifNAT@numbers\if@filesw\immediate\write\@auxout{\string\global\string\NAT@numberstrue}\fi\fi}{}}
\makeatother
\usepackage{amsmath}

\begin{document}


\title{High-resolution Imaging of the Gegenschein \\and the Geometric
Albedo of Interplanetary Dust}


\author{Masateru \textsc{Ishiguro}}
\affil{Astronomy Program, Department of Physics and Astronomy, Seoul
National University, \\599 Gwanak-ro, Gwanak-gu, Seoul 151-742, Republic of
Korea}
\email{ishiguro@astro.snu.ac.kr}

\author{Hongu \textsc{Yang}}
\affil{Astronomy Program, Department of Physics and Astronomy, Seoul
National University, \\599 Gwanak-ro, Gwanak-gu, Seoul 151-742, Republic of
Korea}

\author{Fumihiko \textsc{Usui}}
\affil{Department of Infrared Astrophysics, Institute of Space and
Astronautical Science (ISAS), JAXA, \\ 3-1-1 Yoshinodai, Chuo-ku, Sagamihara,
Kanagawa 252-5210, Japan }

\author{Jeonghyun \textsc{Pyo}}
\affil{Korea Astronomy and Space Science Institute (KASI), Daejeon
305-348, Republic of Korea}

\author{Munetaka \textsc{Ueno}}
\affil{Center for Space Science Technology, Institute of Space and
Astronautical Science (ISAS), JAXA, \\ 3-1-1 Yoshinodai, Chuo-ku, Sagamihara,
Kanagawa 252-5210, Japan }

\author{Takafumi \textsc{Ootsubo}}
\affil{Astronomical Institute, Tohoku University, 6-3 Aramaki, Aoba-ku,
Sendai 980-8578, Japan}

\author{Suk Minn \textsc{Kwon}}
\affil{Department of Science Education, Kangwon National University,\\
192-1 Hyoja-dong, Kangwon-do, Chunchon 200-701, Republic of Korea}

\author{Tadashi \textsc{Mukai}}
\affil{Department of Earth and Planetary Sciences, Kobe University, \\
1-1 Rokkodai-cho, Nada-ku, Kobe 657-8501, Japan}



\begin{abstract}
We made optical observations of the Gegenschein using a
liquid-nitrogen-cooled wide-field camera, Wide-field Imager of Zodiacal
light with ARray Detector (WIZARD), between  March 2003 and November
2006. We found  a narrow brightness enhancement 
superimposed on the smooth gradient of the Gegenschein at the exact
position of the antisolar point. Whereas the Gegenschein morphology
changed according to the orbital motion of the Earth, the maximum
brightness coincided with the antisolar direction throughout the year.
We compared the observed morphology of the Gegenschein with those of
models in which the spatial density of the interplanetary dust
cloud was considered and found that the volume scattering phase function
had a narrow 
backscattering enhancement. The morphology was
reproducible with a spatial distribution model for infrared zodiacal
emission. It is likely that the zero-phase peak (the so-called opposition
effect) was caused by coherent backscattering and/or shadow-hiding
effects on the rough surfaces of individual dust  particles. These
results suggest that big particles are responsible for both zodiacal
light and zodiacal emission.
Finally, we derived the geometric albedo of the smooth
component of interplanetary dust, assuming big particles,
and obtained a geometric albedo of 0.06$\pm$0.01. The derived albedo
is in accordance with collected dark micrometeorites and observed
cometary dust particles. We concluded that chondritic particles are
dominant near Earth space, supporting the recent theoretical study by
dynamical simulation.
\end{abstract}

\keywords{interplanetary medium --- instrumentation: miscellaneous ---
comets: general --- minor planets, asteroids}

\section{Introduction}

The Gegenschein is a faint glow in the sky in the antisolar direction.
Several hypotheses had been suggested to explain the Gegenschein:
sunlight reflected by particles concentrated near the libration point in
the Sun--Earth--particles system, emission from the Earth's gaseous tail,
and sunlight scattered by the Earth's dust tail. After the detection of
the Gegenschein at 1.86 AU from the Sun by the Pioneer 10 spacecraft
\citep{weinberg1973,hanner1974}, it has been widely accepted that the
Gegenschein is a portion of zodiacal light enhanced by the
backscattering of interplanetary dust particles.
Intensive observations of the Gegenschein date back nearly half a
century. It was observed by integrating the brightness in the
field of view using photomultipliers attached to telescopes or
telephotographic lenses
\citep{tanabe1965,roosen1970a,dumont1975,kwon2004}.  These authors observed
the sky background within a certain interval on the sky (1\arcdeg--5\arcdeg) or
scanned the sky with large-field-of-view instruments.
A volume scattering phase function of the interplanetary dust particles was
deduced based on  low-spatial-resolution data
\citep{dumont1975,leinert1976,hong1985,lumme1985,lamy1986}.

Modern observation techniques enable us to obtain snapshot images of the
Gegenschein. \citet{james1997} were the first to used a cooled CCD camera
attached to a fish-eye lens for the observation of zodiacal light and the Gegenschein, obtaining  sufficient resolution (4.8\arcmin~
pixel$^{-1}$) images. Subsequently, a research group at Kobe
University obtained images using a Peltier-cooled CCD
camera (with a quantum efficiency of 10\% at 0.44 \micron) with an
equatorial mounting and succeeded in detecting asteroidal
dust bands on the Gegenschein \citep{ishiguro1999a,ishiguro1999b}.
Note, however, that they applied 1\arcdeg $\times$
1\arcdeg~ boxcar smoothing to reduce the pixel-to-pixel noise level,
which could blur image resolution in many cases. The recent space
observation with the Solar Mass Ejection Imager (SMEI) provided nearly
all-sky photometric maps for over 5 years with a moderate spatial
resolution of 30\arcmin~ \citep{buffington2009}. Note that
\cite{buffington2009} stacked data over the mission period because the
S/N ratio was inadequate to produce the Gegenschein map. They did not
consider the seasonal variation in the Gegenschein morphology.

The first question we would like to address in this paper is the exact
position of the maximum brightness in the
Gegenschein. \citet{dumont1965} and \citet{roosen1970a,roosen1970b}
found that the maximum brightness was close to the antisolar point.
However, \citet{tanabe1965} and \citet{wolstencroft1967} reported that
the maximum brightness of the Gegenschein was displaced from the
antisolar point and existed close to the invariable plane of the Solar
System. \citet{mukai2003} argued that the maximum brightness was largely 
influenced by the existence of fine-scale structures in the zodiacal
cloud, i.e., asteroidal dust bands
\citep{sykes1986,sykes1990,reach1997}, based on their CCD
observation. After  subtraction of the fine structures, they
estimated a deviation from the antisolar point of $-$0.4\arcdeg~ from
an image taken on 2 November 1997. \citet{buffington2009} found that the
maximum existed at the antisolar point,
although the study was based on SMEI measurements, whose spatial
resolution (0.5\arcdeg) is equivalent to the displacement discussed in
\citet{mukai2003} and \citet{ishiguro1998}.
High-resolution observations ($<$0.5\arcdeg) are thus essential to
settle the matter of the maximum brightness position.

The second question concerns the Gegenschein morphology. A systematic
morphological
change was pointed out by \citet{tanabe1965} and later confirmed by
high-resolution observations \citep{james1997,ishiguro1998,kwon2004}.
This change can be caused by the discrepancy between the orbital plane of the
Earth (i.e., the ecliptic plane) and the plane of symmetry of the zodiacal
cloud. The spatial distribution of the zodiacal cloud has been
investigated around the solar elongation $\epsilon$ $\sim$
90\arcdeg~ \citep{kelsall1998}.  However, little is known about the
spatial distribution around the Gegenschein region, which might retain
information about the spatial distribution of  interplanetary dust
particles that exist further than those observed around $\epsilon$
$\sim$90\arcdeg.

In this paper, we present high-resolution images of the
Gegenschein taken by a wide-field CCD camera for the diffuse night sky,
 Wide-field Imager of Zodiacal light with ARray Detector (WIZARD).  Both
the sensitivity and the spatial resolution are better than those of the
instruments used previously
\citep{james1997,ishiguro1998,mukai2003,buffington2009,kwon2004}.
We examined the morphology and the photometric center of the Gegenschein
for 4 years using WIZARD. We succeeded in  detecting the narrow brightness enhancement at
the exact position of the antisolar direction. The maximum brightness
of the Gegenschein consistently existed at the antisolar point throughout
our observation period. We examined the volume scattering phase function
around the opposition and considered the reason for the narrow
brightness enhancement. In Section 2, we describe the observations and
the data reduction. In Section 3, we show the morphological maps of the
Gegenschein. In addition, we discuss the annual variation in the peak
position. In Section 4, we test several phenomenological three-dimensional
models to interpret the observational results. Finally, we discuss the
geometric albedo of  interplanetary dust particles in the last
section.

\section{Observations}

\subsection{The Instrument and its Performance}

All data in this paper were taken with WIZARD. A detailed description of
the instrument is given in \citet{usui2002}, \citet{ishiguro2003}, and
\citet{ueno2007}.
In this section, we briefly describe the instrument and 
report on its performance based on the data reduction results.

\begin{figure}
 \epsscale{0.75}
   \plotone{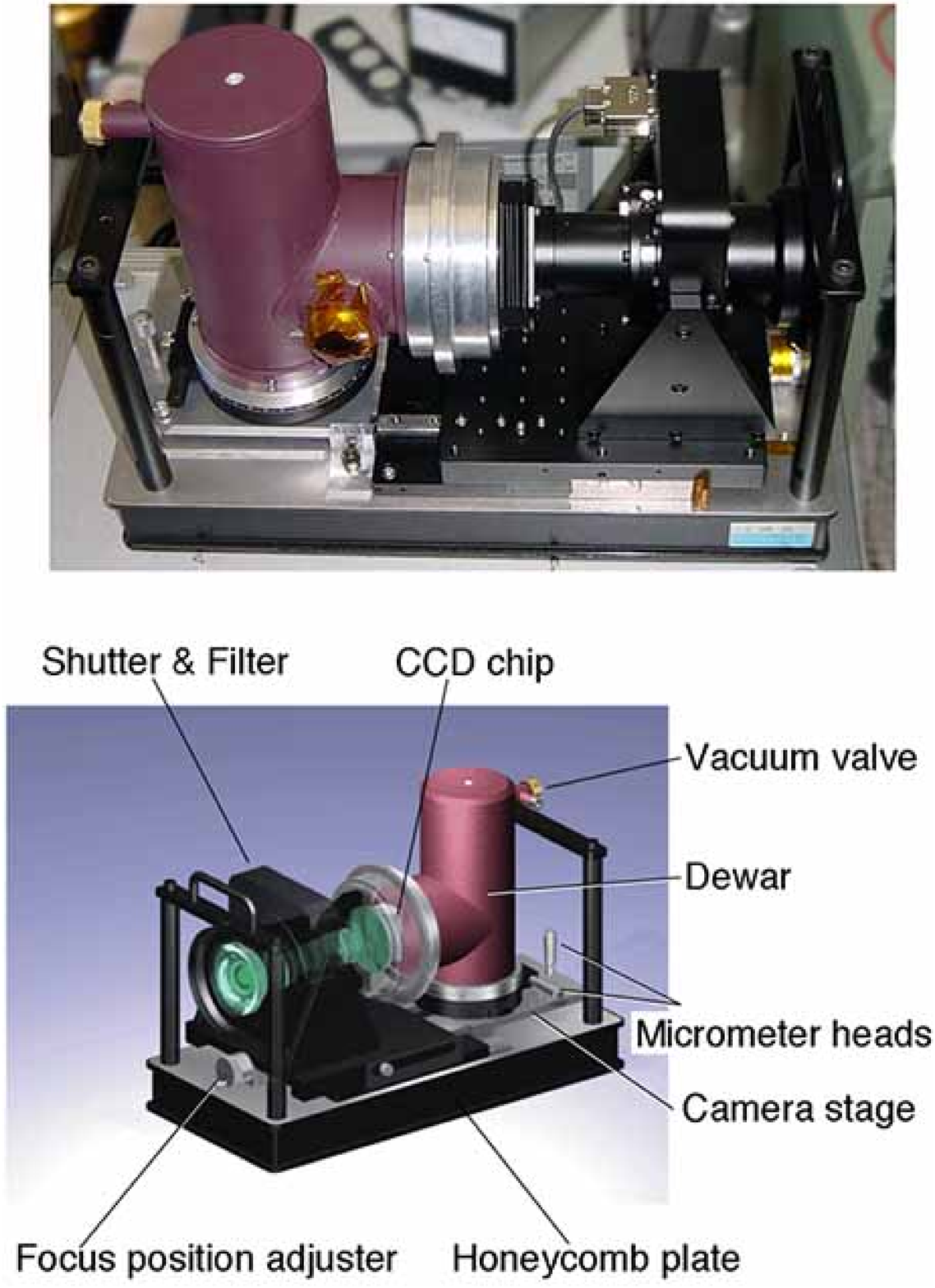}
  \caption{Photograph (\textit{top}) and schematic view (\textit{bottom}) of WIZARD. It
 consists of a cryogenic dewar and  optics. These two units are
 connected with a bellows tube and mounted on a thick honeycomb plate.
 The mechanical shutter  and  glass
 filter are placed at the pupil position. The system is mounted on an
 equatorial mounting for star-tracking.}\label{fig:wizard-photo}
\end{figure}

\begin{figure}
 \epsscale{1.0}
   \plotone{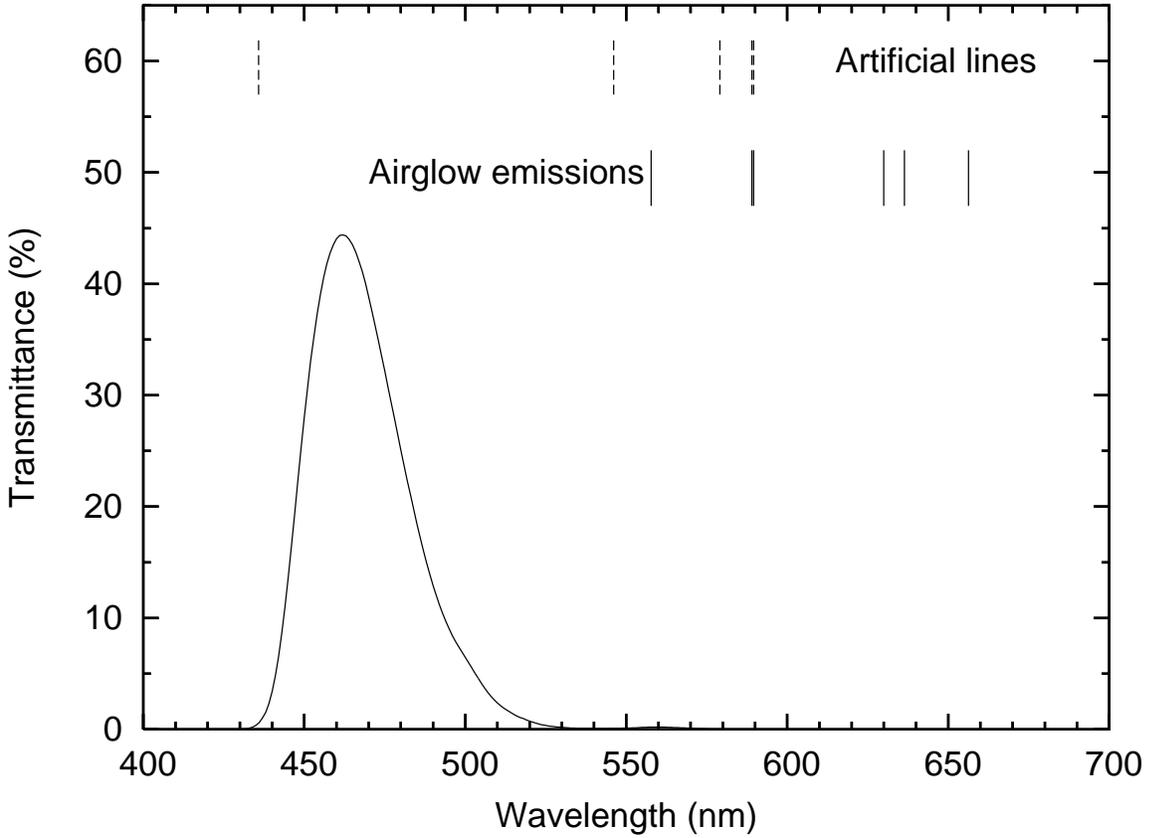}
  \caption{Transmittance of the filter we applied for the Gegenschein
 and zodiacal light observations. This filter is  designed to avoid
 the prominent  natural emissions and artificial emissions, indicated by
 solid lines and dashed lines, respectively. They consist of
 \ion{O}{1} (557.7, 630.0, and 636.4 nm), NaI (589.0 and 589.6 nm), and
 H$\alpha$ (656.3 nm) for the natural emissions and
 HgI (435.8, 546.1, and 579.1 nm) and NaI (589.0 and 589.6 nm)
 for the artificial emissions, respectively.
}\label{fig:transmittance}
\end{figure}

WIZARD was developed to cover a wide-extended  zodiacal light and
the Gegenschein. It consists of a camera head and a wide-field lens (Figure
\ref{fig:wizard-photo}). The system was mounted on an equatorial
mounting. We employed a back-illuminated full-frame CCD chip, EEV CCD
42-80 ($27.6\times 55.3$ mm; 2048$\times$4096 pixels), placed in a compact
N$_2$-cooled cryogenic dewar (Kadel Engineering Corp.). During
observations, the CCD temperature was maintained at around 180 K. The optical
system was designed  by the {Genesia}
Corporation. It has a large image circle (60 mm in diameter) to cover
the whole imaging area of the CCD. The focal ratio and the focal length
are 2.8 and 32.5 mm, respectively.  With the camera head, it gives a
field of view of $46\arcdeg\times 92\arcdeg$ with a pixel resolution
of 1.4\arcmin. Figure \ref{fig:transmittance} shows the transmittance of
the optical filter used for zodiacal light observations. It covers a
wavelength range of 440--520 nm, avoiding prominent airglow
emissions and artificial emission light. In particular, it was designed
not to transmit \ion{O}{1} airglow at 557.7 nm and artificial mercury vapor
light at 435.8 nm. In addition, sodium beacons at 589.2 nm as laser guide
stars are nowadays problematic for astronomical observations.
The transmittances at these wavelengths are negligible, that is, 0.35\%
at 435.8 nm (Hg), 0.19\% at 557.7 nm \ion{O}{1}, and 0.0008\% at 589.2 nm (Na),
respectively.  The CCD clock pattern and the operation commands were
prepared on a Linux host computer and sent to the CCD chip through a
COGITO-3 system. COGITO-3 is a programmable control system
for  imaging devices specifically designed for scientific purposes and was developed by
Ueno and Wada at ISAS/JAXA. It contains three units: a sequencer board,
an A/D board, and a driver board.  The obtained analog signal was
converted to the digital one on the COGITO-3 A/D board and transferred to
the host computer via a PCI--VME bus interface adapter.

The analog signal in each pixel was sampled 16 times to reduce the
readout noise. In addition, during CCD
readout, we turned off the pulse motor used for star-tracking to reduce noise associated with the motor drive unit. Consequently, the readout
noise was reduced to the level of 20--25 e$^-$.  To monitor the
zero level of the CCD, we recorded the overscan data by reading 50 dummy
pixels before and after each row was read.
For confirmation, we examined  the stability of
the zero intensity  level using dark frames obtained at the intervals of the
light frame. We found that the standard deviation of the zero level in
each frame was $\lesssim$10 e$^-$. Using the overscan region, we
determined the zero level to an accuracy of  4 e$^-$. The typical
brightness of the Gegenschein plus airglow was about 1.1 e$^-$
s$^{-1}$. With an exposure time of 600 s, the  uncertainty of
the zero intensity  level (i.e., 4 e$^-$) was small enough (0.6\% of the
Gegenschein plus airglow brightness, i.e., 660 e$^-$) to derive the
brightness of the Gegenschein plus airglow. In addition,  background
noise becomes dominated by 
photon noise from the natural night sky background rather
than the readout noise. For this reason, we fixed the exposure time of
600 s for our Gegenschein observations.

\begin{figure}
 \epsscale{1.0}
   \plotone{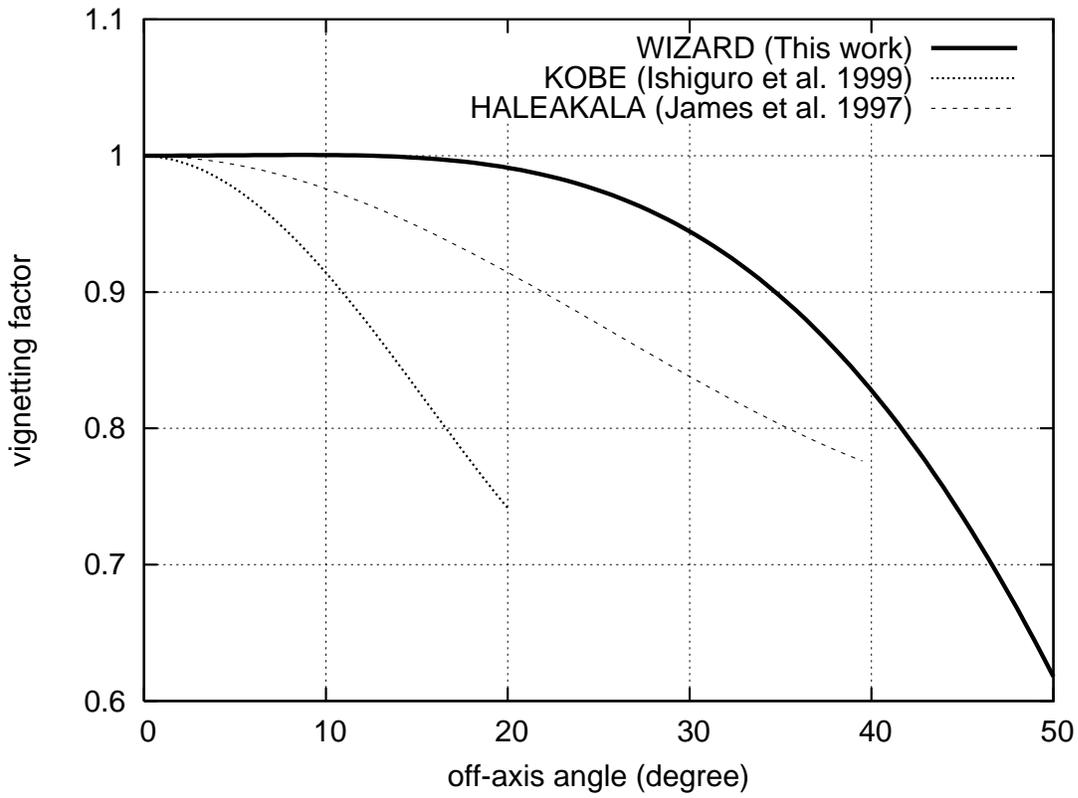}
  \caption{Vignetting factor of WIZARD normalized at the optical center
 (\textit{thick solid line}). Two vignetting functions in \citet{ishiguro1999a}
 and \citet{james1997} are also drawn for comparison. It is clear that
 WIZARD optics has little vignetting within 20\arcdeg~ from the
 center. }\label{fig:flat}
\end{figure}

An image taken by a wide-field camera generally shows strong
vignetting. The detected intensity around the optical center is thus
brighter than that at the edge of the frame. A flat-field correction is
essential to obtain  accurate intensity maps of the zodiacal light
and the Gegenschein. It is, however, difficult to obtain flat data using
WIZARD because of its huge field of view. We investigated
flat-fielding for different spatial scales, that is,  by considering optical
vignetting as a low-frequency component and  pixel-to-pixel variation
in the sensitivity as a high-frequency component. We examined
the optical vignetting by observing bright stars at  different
positions on the CCD chip. We confirmed that the weather
conditions were stable and the airmasses of these stars were constant during
the exposures, suggesting that the brightnesses of these light sources
were stable enough to examine the vignetting. Figure \ref{fig:flat}
shows the vignetting function across
the CCD chip. From the measurement, we found that the detected intensity
was almost constant within 20$^{\circ}$ from the optical center and
reduced by 15\% at 38$^{\circ}$  and 30\% at 47$^{\circ}$. For
reference, we compare the vignetting function with 
those of previous CCD observations of the Gegenschein
\citep{james1997,ishiguro1999a}, in which  commercial products for
single-lens reflex cameras were used. It is clear that the optics for WIZARD is
well designed to suppress the vignetting throughout a large portion of the CCD
chip. Small-scale (pixel-to-pixel) fluctuations of the CCD sensitivity
were examined by taking images with a Labsphere integrating sphere.
We found that the pixel-to-pixel responsivity variation was negligibly
small ($\lesssim$0.3\%).

The performance of WIZARD, which are critical for estimating the error
in the measurements, are summarized in Table 1.

\subsection{Observations and Data Analysis}
\label{sec:analysis}

In normal operation, we began observation soon after the end of
evening twilight. We took a couple of images successively without
slewing the camera to its next position. Since the motor controller for star-tracking was turned off for about 3 min after exposures for the CCD
readout, the position of celestial objects moved about one degree on the
CCD chip. This observation strategy enables us to distinguish small-scale
structures of the Gegenschein from instrumental artifacts associated
with uneven sensitivity of the instrument.

\begin{figure}
 \epsscale{1.3}
   \plotone{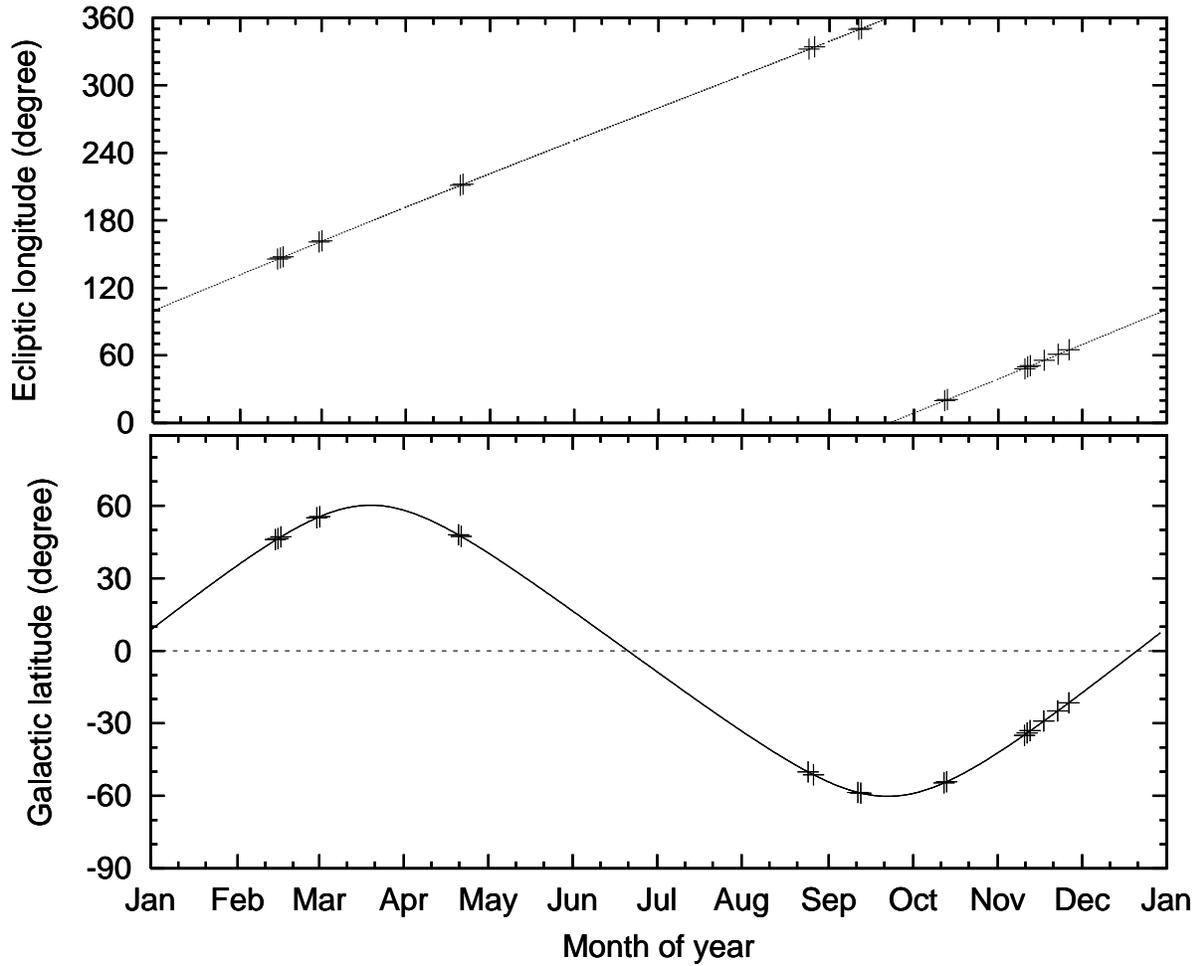}
  \caption{Ecliptic longitude (\textit{top}) and Galactic latitude (\textit{bottom}) of the
 antisolar point as a function of the day of the year. Crosses denote the
 ecliptic longitude and Galactic latitude when we made the
 observations. The antisolar point lies on the Galactic plane on June
 21 and December 22.}\label{fig:condition}
\end{figure}

We made a sequence of observations of the zodiacal light and the
Gegenschein with
WIZARD between March 2003  and  November 2006 on Mauna Kea. Our
observations were supported by the Subaru telescope (National
Astronomical Observatory of Japan) and the InfraRed Telescope Facility (ITRF; NASA and Institute for
Astronomy, University of Hawaii). We observed zodiacal light and the
Gegenschein for 45 nights under  photometric
conditions. We selected Gegenschein data taken at high Galactic latitude
near the zenith. In total, 14 nights of data in  different seasons of the
year were analyzed for the present study.  We summarize the observation
conditions of these data in Table 2.  Gegenschein data were
not available in May--July and December--January because the
antisolar point was close to the Galactic plane (Figure \ref{fig:condition}).

The night sky brightness observed from the ground consists of those of
zodiacal light including the Gegenschein, integrated starlight of
unresolved stars, diffuse Galactic light, and pseudo-continuum airglow
\citep{leinert1998}. These light sources are attenuated by scattering of
atmospheric molecules and aerosols. At the same time, the scattered
component in  Earth's atmosphere results in a diffuse light
source. \citet{dumont1965} introduced the concept of effective optical
depth for diffuse light sources, which is different from that for point
sources. Later,
\citet{hong1998} and \citet{hong2002} developed the idea by solving
radiative transfer in an anisotropically scattering spherical atmosphere
of the Earth. It is difficult reduce zodiacal light data in high-airmass (i.e., large zenith distance) regions owing to the considerable
contributions of the atmospheric diffuse light and the strong zenith
extinction, although some intensive studies were conducted to obtain the
absolute brightness of the zodiacal light from ground-based
observatories \citep{kwon2004,dumont1965}. The data reduction of the
Gegenschein is, however, relatively easy. The
Gegenschein appears near the zenith
around  midnight on Mauna Kea, where the contaminations from  airglow
and  zenith extinction are smallest in the sky. The typical
brightness of atmospheric diffuse light components was 20
S$_{\odot10}$ and had a weak slope of 10\%--20\% of the Gegenschein
brightness over a 40\arcdeg~ field of view in Gegenschein
exposures, where 1 S$_{\odot10}$ is equivalent to the flux of a solar-type star of magnitude 10 distributed over one degree squared
\citep{leinert1998}. Inadequate reduction for these atmospheric
components does not tend to lead to a misunderstanding of  the morphology of the
Gegenschein. In addition to these diffuse light sources mentioned above, there
remain a number of point sources. Stars brighter than a visual magnitude
of 12--14 (depending on the Galactic latitude) were resolved by our
observations. We subtracted point sources brighter than  magnitude 12
using the USNO--B2.0  catalog \citep{monet03}.
The detected pixel data were replaced by the
median values of circumjacent pixels.

\begin{figure}
 \epsscale{1.0}
   \plotone{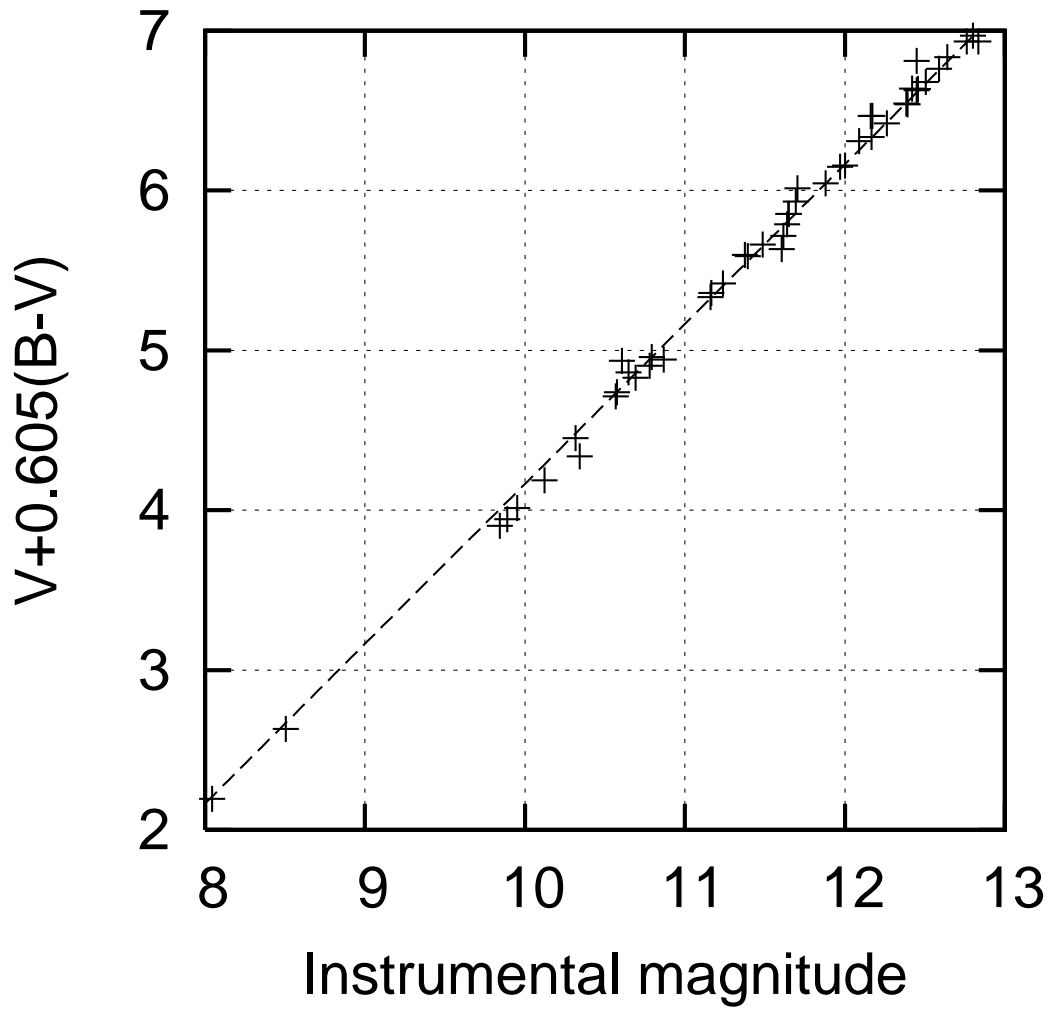}
 \caption{Comparison between the instrumental magnitude defined as $-2.5\log$(DN) + constant and color-corrected catalog magnitude of field
 stars, where $DN$ denotes CCD count. Since these stars exist near the
 zenith (within an 
 airmass of 1.07), we did not correct for the zenith extinction term (see
 Section \ref{sec:analysis}).
 }\label{fig:color-mag}

\end{figure}

For the most part, the data were analyzed as follows:
First, the
raw data were reduced using the overscan region as a bias level plus
flat-field data. 
The flux calibration factor 
and the optical depth of the point sources are determined by
aperture photometry of field stars at various
airmasses.  Since we applied a nonstandard filter, we constructed a
photometric standard star catalog for WIZARD data. An empirical formula
is applied to convert from the Johnson-Cousins system into the WIZARD
system, that is, $M_{WZD}$ = $V + \alpha (B-V)$, where
$\alpha$ is the color conversion factor and $B$ and $V$ are B-band and
V-band magnitudes, respectively. To derive the coefficient $\alpha$, we selected 56 stars
in BSC Ver. 5 in an airmass range of 1.00 to 1.07. Since the typical
extinction coefficient on Mauna Kea in the WIZARD band was $\sim\negthickspace0.15$,
atmospheric extinction did not change the stellar magnitude by $\gtrsim$0.01
magnitude in this sky area, which is adequate to derive the
coefficient $\alpha$.  The result is shown in Figure
\ref{fig:color-mag}. In the figure, the bottom axis denotes the
instrumental magnitude measured by  aperture photometry, and the
perpendicular axis denotes the magnitude in the WIZARD system ($M_{WZD}$). The best-fit parameter was found to be  $\alpha=0.605\pm 0.005$.

Unresolved starlight brightness plus diffuse Galactic light brightness
were subtracted by using Pioneer 10 and 11 data, which was obtained
beyond the inner 
Solar System (i.e., outside the zodiacal cloud) and the USNO--B2.0  catalog
\citep{monet03}.  Although the central wavelength of WIZARD (460 nm) is
slightly longer than that of the Pioneer blue band (440 nm), we used the
Pioneer data without color correction.
The contribution from unresolved starlight (stars fainter than
magnitude 12) was estimated to be 20 S$_{\odot10}$ (or 10\% of the Gegenschein
brightness) at a Galactic latitude of 25\arcdeg~ and  10 S$_{\odot10}$
(or 5\% of the Gegenschein brightness) at a Galactic latitude of
50\arcdeg~ \citep{leinert1998}.
The intensity of the airglow was subtracted using
Barbier's function \citep{barbier1944,barbier1961}, whereas the intensity
of the multiple-scattering component in Earth's troposphere was not
subtracted because of its weak contribution in our data around the
Gegenschein. The
extinction coefficients of  Earth's atmosphere were obtained through
 stellar photometry, while the effective optical depths for the
extended sources were obtained using the QDM\_sca code by adjusting the particle
albedo $\omega$ and the asymmetry factor $g$
\citep{hong1998,hong2002} to match the brightness of previous zodiacal
light observations \citep{levasseur1980}.
Some of unknown parameters such as the intensity of the
airglow at zenith and scaling factor of starlight brightness by Pioneer
to our system were determined empirically to fit our data to the zodiacal
light brightness in \citet{levasseur1980}. We derived the Gegenschein
brightness in the range of 165--190 S$_{\odot10}$.
Throughout this data reduction, we found that there is an uncertainty of $\pm$10--15
S$_{\odot10}$ in the absolute brightness. This uncertainty is mainly
caused by inadequate subtraction of atmospheric diffuse light sources
and imperfect correction of the zenith extinction, although instrumental effects
such as scattered light inside the camera cannot be ruled out.
It is, however, emphasized that the relative intensity distribution of
$\lesssim$20\arcdeg~ structures is reliable in our data set because it
cannot be influenced by 
the inevitable atmospheric and instrumental effects.

\section{Results}

\subsection{Observational Results}

\begin{figure}
 \epsscale{1.0}
   \plotone{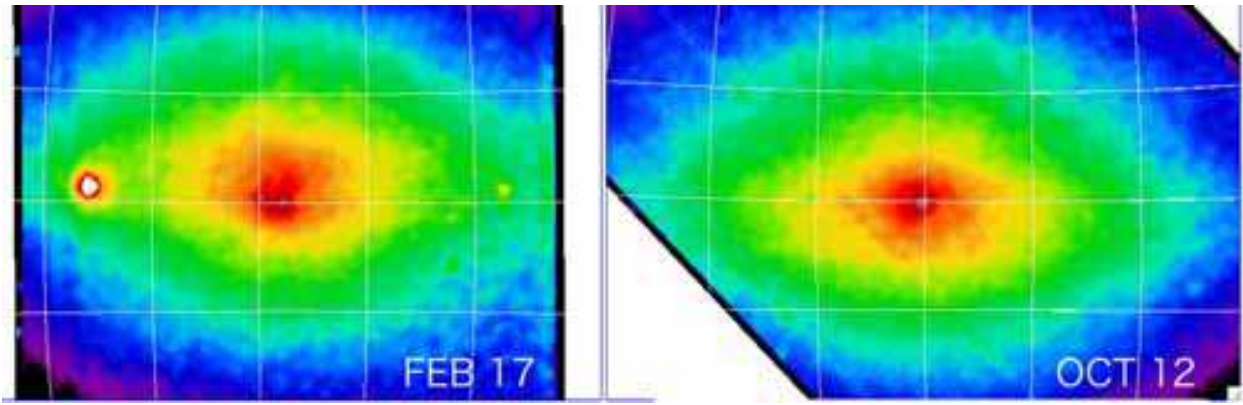}
  \caption{Contour maps of the Gegenschein taken on (\textit{left}) February
 17 and (\textit{right}) October 12. The interval of grid is 10\arcdeg.}\label{fig:contour}
\end{figure}

Figure \ref{fig:contour} shows  snapshots of the Gegenschein in
ecliptic coordinates. The long axis was nearly parallel to the ecliptic
and stretched 
in the east--west direction. The morphology was symmetrical with
respect to the line at an ecliptic longitude with respect to the Sun
of $\lambda - \lambda_{\odot}$ = 180\arcdeg, where $\lambda$ and
$\lambda_{\odot}$ denote the ecliptic longitude and ecliptic longitude
of the Sun. Looking at the morphologies closely, we see  a truncated oval
shape whose eastern and western edges are angular rather than round. As
the antisolar point is approached, the shape changes from truncated
oval to round. There is a prominent peak at the antisolar point.

It is clear that the brightness distribution shifted
northward on 17 February  and  southward on 12 October  (Figure
\ref{fig:contour}). We found that the Gegenschein morphology deviated
to the south in August--October and to the north in
February--April. This trend is consistent with the result of
\citet{ishiguro1998}, who mentioned that the overall brightness
distribution shifted  southward in September but shifted
northward in March. The annual variation can be explained by the
three-dimensional zodiacal cloud model having a plane of symmetry deviated
 southward during August and November and  northward during
February and April, as previously noticed in \citet{ishiguro1998}. We
discuss the annual variation in Section 3.2.

\begin{figure}
 \epsscale{1.0}
   \plotone{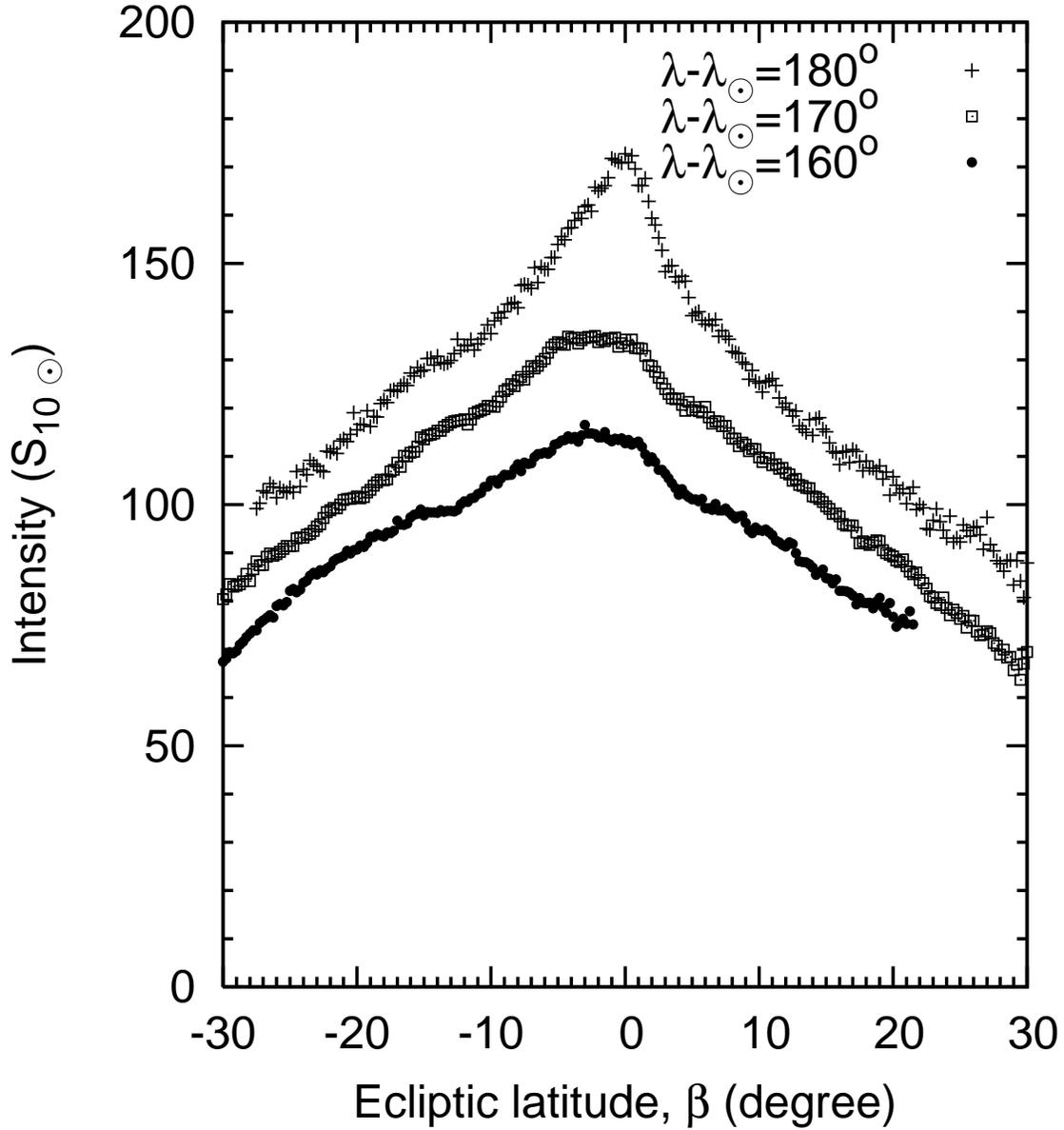}
 \caption{Surface brightness profile along the ecliptic latitude at
 $\lambda - \lambda_{\odot}=180\arcdeg$ (\textit{top}), 170\arcdeg~(\textit{middle}), and
 160\arcdeg~(\textit{bottom}). For clarity, the middle and lower profiles are
 shifted by 10 S$_{\odot10}$ and 20 S$_{\odot10}$,
 respectively.}\label{fig:lat_profile}
\end{figure}

Figure \ref{fig:lat_profile} shows
the surface brightness profile of the Gegenschein on 12 October at three
different longitudes. These profiles show a convex feature over
the ecliptic latitude range $\beta=-3\arcdeg$ to $1\arcdeg$. The
flat-topped profiles were detected for the first time through the infrared
observation of zodiacal emission \citep{low1984}. They were also
recorded for the Gegenschein \citep{ishiguro1999a,mukai2003}. Recent
theoretical studies have showed that features in the profiles are
associated with  materials generated by  recent catastrophic collisions
in the main-belt asteroid region
\citep{nesvorny2003,nesvorny2006}. However, the convex structure is not
clear in the profile of $\lambda-\lambda_{\odot}$ = 180\arcdeg. Instead,
it exhibits a steep increase at $\beta=0\arcdeg$.

\begin{figure}
 \epsscale{0.6}
   \plotone{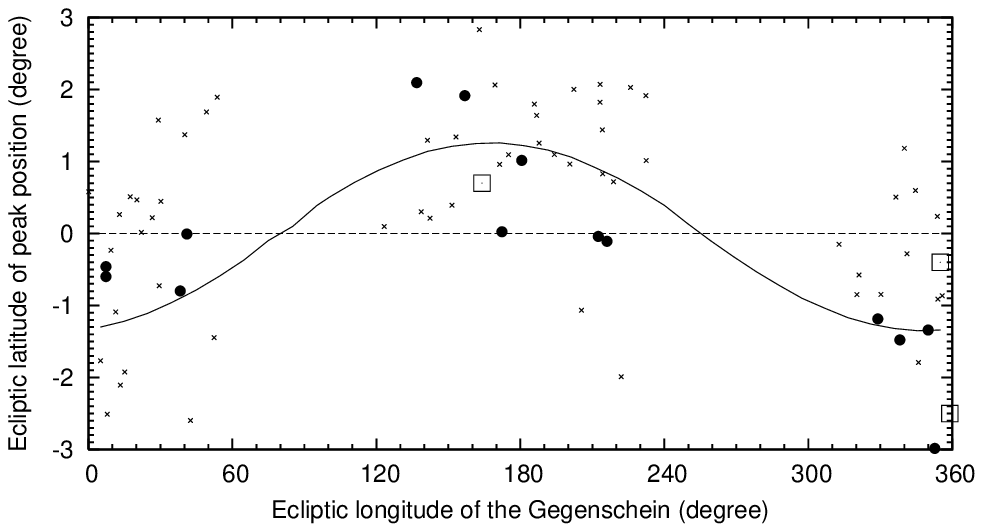}
   \plotone{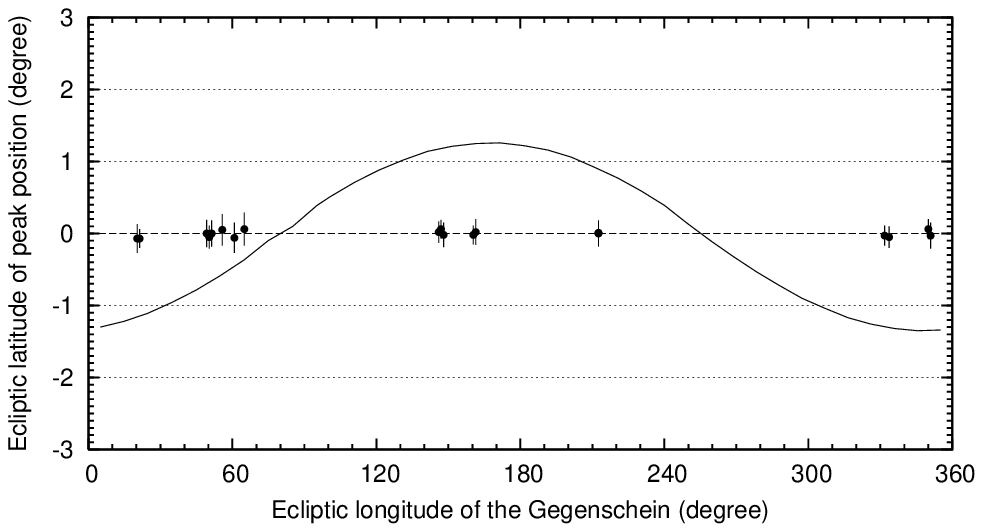}
\caption{Observed latitude of the maximum intensity of the Gegenschein.
In the top panel, the data are from photographic (\textit{crosses}),
 photoelectric (\textit{filled circles}), and CCD observations (\textit{open
 squares}) \citep{roosen1970b,ishiguro1998}.
 Our results are shown in the bottom panel. For comparison, we show the
 position predicted using a plausible model on the basis of
 \citet{mukai2003} and \citet{kwon2004} (\textit{solid line}). In the model, the
 plane of symmetry for $i_\mathrm{sym}=2\arcdeg$ and
 $\Omega_\mathrm{sym}=80\arcdeg$ is
 considered.  The scattering phase function in \citet{hong1985} was
 applied for the  model, assuming $\nu=1.3$.}
\label{fig:peak-position}
\end{figure}

Despite the annual variation in the overall shape, we found that the
maximum position coincided with the antisolar point through the
observation runs.
Neither east--west nor north--south displacement was detected in the
profiles at $\lambda - \lambda_{\odot} =180\arcdeg$.
Figure \ref{fig:peak-position} shows the ecliptic latitude of the
maximum intensity at $\lambda - \lambda_{\odot} =180\arcdeg$.
For comparison, we show photoelectric and photographic data
compiled by \citet{roosen1970b} as well as  previous CCD data.
Our results are inconsistent with those in \citet{roosen1970b},
\citet{james1997}, \citet{ishiguro1998}, and \citet{mukai2003}, but they are
consistent with the recent measurement from space by SMEI
\citep{buffington2009}. It should be emphasized that our measurement is the
most reliable of all observational data so far because of the better
spatial resolution.

\subsection{Comparison with 3-D Empirical Dust Distribution Models}

In this chapter, we analyze the annual variation in the morphology and
the peak position by conducting  three-dimensional
zodiacal cloud model simulations.

The surface brightness of the zodiacal light as seen by an observer in
the ecliptic plane at an ecliptic latitude $\beta$ and an ecliptic longitude
$\lambda-\lambda_{\odot}$ with respect to the Sun
can be described as a double integral over the size distribution and
along the line of sight. As an approximation, we assume the following:\\

\begin{enumerate}
\item The size distribution of dust particles in the zodiacal cloud is
    independent of their position in  interplanetary space.

\item The albedo and scattering phase functions of dust particles are
      independent of  position.

\item The zodiacal cloud has a plane of symmetry and is axisymmetric with
      respect to an axis perpendicular to the plane of symmetry.

\item Dust grains are sufficiently far from the Sun that we can
    regard the Sun as a point source.
\end{enumerate}

\begin{figure}
 \epsscale{1.0}
   \plotone{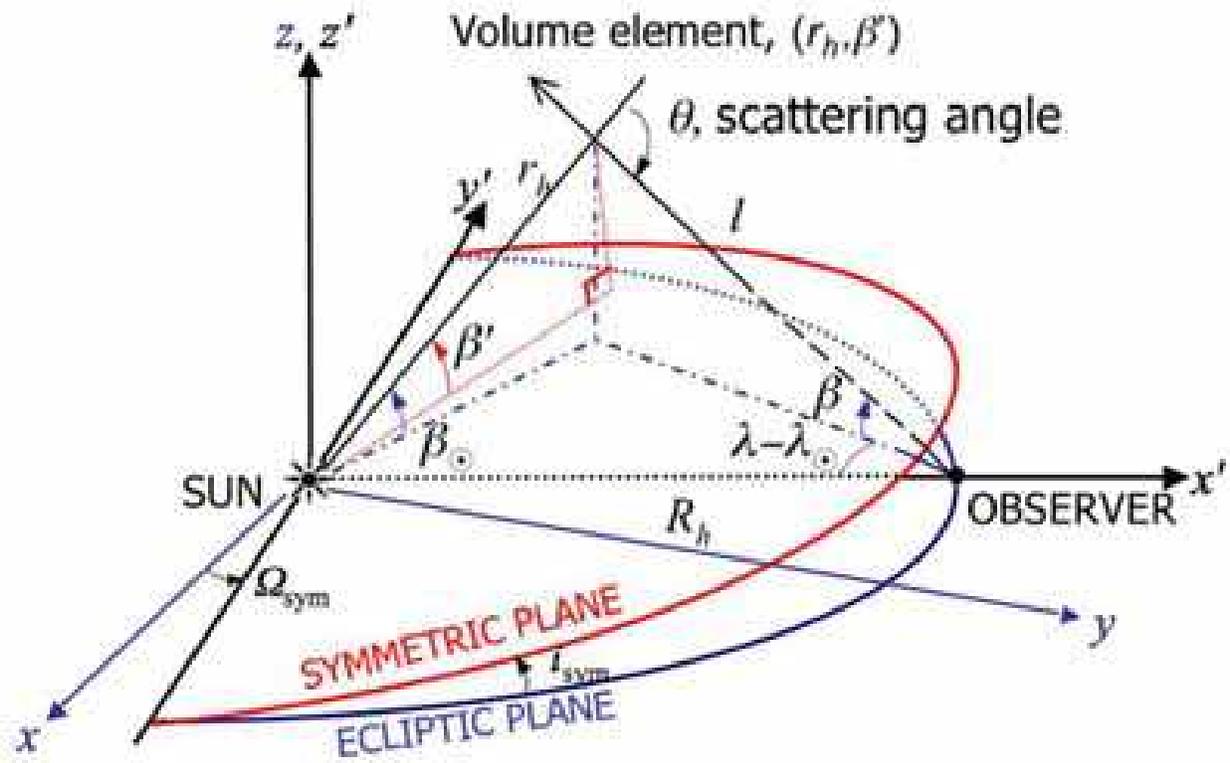}
\caption{The viewing geometry for light scattered by a dust volume
 element. The scattering angle $\theta$ is measured from the
 line of sight to the antisolar direction at the volume element.}
\label{fig:geometry}
\end{figure}

\noindent
Under these assumptions, the brightness of the zodiacal light is given by
\begin{eqnarray}
I_{ZL}(\lambda-\lambda_{\odot}, \beta)=
\int_{0}^{\infty} \int_{0}^{\infty}
F(r_h) n(x,y,z) \,f(s)\,\sigma_{sca}(s)\,\phi(\theta; s) \,ds\,dl,
\end{eqnarray}

\noindent
where $dl$ and $\theta$ denote the line element and the scattering
angle, respectively, and $s$ is the particle radius. $f(s)$ is the
differential size distribution of the particles with a radius of $s$.
$\sigma_{sca}(s)$ is the cross section of the particles.
$\phi(\theta; s)$ characterizes the distribution of scattered intensity
with $\theta$.
The incident solar flux $F(r_h)$ can be replaced by
$F_{\odot}(r_0/r_h)^2$, where $F_{\odot}$ denotes the solar flux at the
heliocentric distance $r_h=r_0$.  $n(x,y,z)$ is the number density of
the cloud particles in ecliptic coordinates with respect to the Sun,
where the $x$ axis points from the Sun toward the vernal equinox, the
$z$ axis is perpendicular to the ecliptic plane, and the $y$ axis completes
a right-handed orthogonal coordinate system. The geometry and the
notations are summarized in Figure \ref{fig:geometry}.

For simplification, the mean volume scattering phase function
$\Phi(\theta)$ and the mean total scattering cross section
$\bar{\sigma}_{scat}$ are defined by

\begin{equation}
\int_{0}^{\infty} f(s)\,\sigma_{scat}(s)\,\phi(\theta;s)\,ds
\equiv \bar{\sigma}_{scat} \,\Phi(\theta),
\end{equation}

\noindent with a normalization condition, that is,

\begin{equation}
\int_{4\pi} \Phi(\theta) \,d\Omega = 1 .
\end{equation}

\begin{figure}
 \epsscale{1.1}
   \plottwo{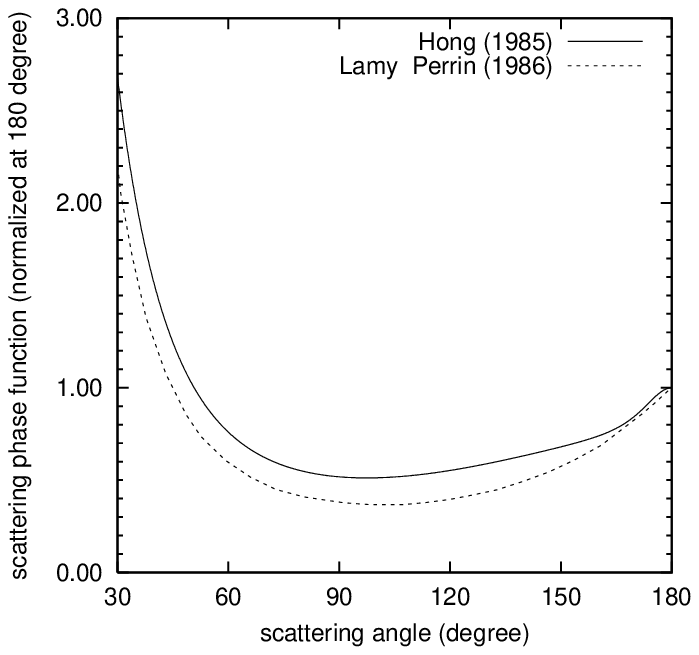}{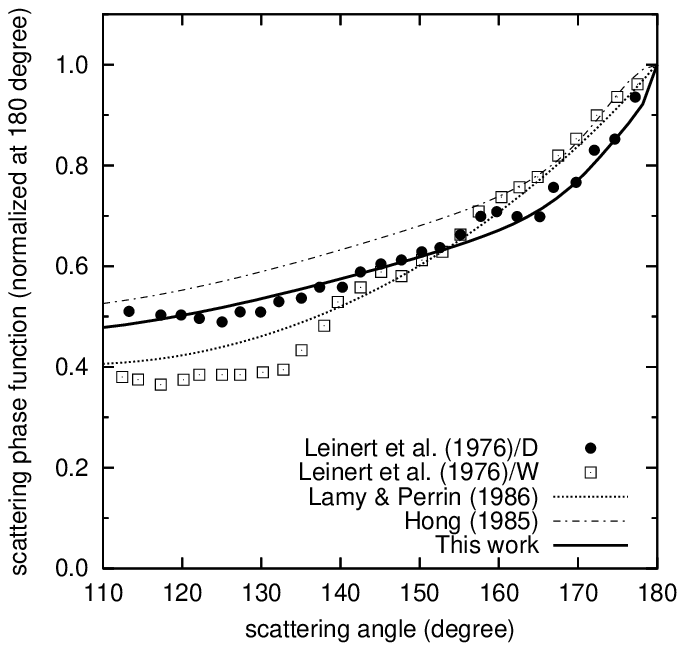}
\caption{(\textit{Left}) The scattering phase functions from \citet{hong1985} and
 \citet{lamy1986} for a radial power law $\nu=1$.
 (\textit{Right}) Comparison between scattering phase functions
 obtained by other authors and us.}
\label{fig:VSF}
\end{figure}

\noindent
The mean volume scattering phase function was empirically derived from the
zodiacal light observations \citep{leinert1976,hong1985,lamy1986}.
Figure \ref{fig:VSF} (left) shows an example of the
mean volume scattering phase function given by \citet{hong1985} and
\citet{lamy1986}. It has a  strong peak in the forward direction,
an isotropic component at
intermediate scattering angle, and a slight enhancement in the backward
direction.

The spatial distribution of dust particles has been discussed
\citep{giese1986,giese1989,kelsall1998}.  
In the early years of research, the spatial distribution was of a form
separable into radial and vertical terms, that is,
\begin{equation}
n(r_h;\beta')=n_0~\left(\frac{r_h}{r_0}\right)^{-\nu} \,h(\beta ') ,
\end{equation}

\noindent
where $n_0$ is the reference dust number density on the symmetry plane
at  heliocentric distance $r_h=r_0$. The radial power law is motivated
by the radial distribution expected for particles under the effect of
Poynting-Robertson drag, which results in $\nu=1$ for dust particles in
circular orbit; $h(\beta ')$ describes how the dust number density 
reduces as we move away from the symmetry plane.
Various axisymmetric
distribution models were proposed. Among them, the ellipsoid model,
\begin{equation}
h(\beta ')=\left[ (1+6.5~\sin~\beta ')^2 \right]^{-0.65}
,\end{equation}
\noindent  and the fan model,
\begin{equation}
h(\beta ')= \exp (-2.1 |\sin \beta '|),
\end{equation}
\noindent were often applied in phenomenological model calculation in the
literature \citep[e.g.,][]{giese1989}. The plane of symmetry was derived
by different authors (see, e.g., \citet{kwon2004}, who obtained $i_\mathrm{sym} \sim
2\arcdeg$ and $\Omega_\mathrm{sym}=80\arcdeg$, and \citet{mukai2003},
who obtained $i_\mathrm{sym}=2.03\arcdeg \pm 0.50\arcdeg$ and
$\Omega_\mathrm{sym}=54\arcdeg$--64\arcdeg). A sophisticated model, in which
 localized dust condensations as well as the axisymmetric
smooth component are considered, was suggested on the basis of infrared observations
with the Cosmic Background Explorer's Diffuse Infrared Background Experiment (COBE/DIRBE) \citep{kelsall1998}. It consists of a  smooth diffuse
component that has a plane of symmetry of
$i_\mathrm{sym}=2.03\arcdeg\pm 0.017$\arcdeg~ and
$\Omega_\mathrm{sym}=77.7\arcdeg\pm 0.6$\arcdeg, three asteroidal dust
components, and circumsolar ring components.

\begin{figure}
 \epsscale{1.0}
\plotone{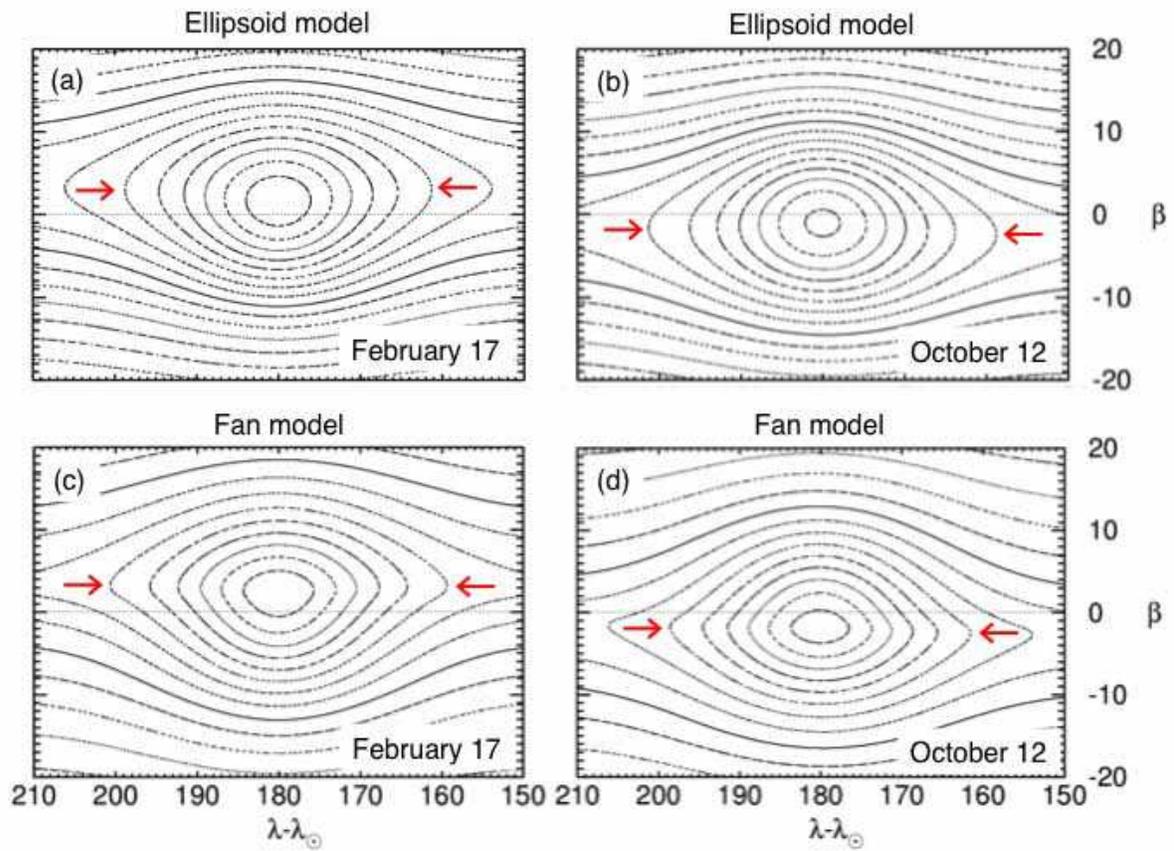}
  \caption{An isophoto map, in arbitrary units, of the Gegenschein at
 two different days of year based on the ellipsoid model (\textit{top}) and the fan
 model (\textit{bottom}). The scattering phase function of \citet{hong1985} was
 applied. }\label{fig:ellfan}
\end{figure}

We began with a comparison of the ellipsoid model and the fan model.
We assumed a plane of symmetry of $i_\mathrm{sym}=2$\arcdeg~ and
$\Omega_\mathrm{sym}=80$\arcdeg~
based on  previous studies \citep{kwon2004,kelsall1998}.
The radial power law exponent $\nu=1.3$ is assumed. We confirmed that
$\nu$ does not change the Gegenschein morphology significantly within
$\sim$30\arcdeg~ from the antisolar point.  The mean volume scattering
phase function in \citet{hong1985} was applied. We should note here that
the power-law exponent $\nu$ is insensitive to the morphology of the
Gegenschein, as already pointed out in \citet{hong1985}. For the reason,
we did not give much attention to $\nu$ in this paper.
The results are shown in Figure \ref{fig:ellfan}. The systematic change
in north--south asymmetry 
is explained qualitatively by these models. The model morphologies are,
however, different from the observed ones in that they have
protruding edges toward the eastern and western directions (see the
arrows in Figure \ref{fig:ellfan}). No peak in the antisolar direction
appears in these modeled images. These mismatches cannot be explained by
different values of $i_\mathrm{sym}$ and $\Omega_\mathrm{sym}$,
suggesting that the classical models are inadequate for explaining the
observed distribution of interplanetary dust particles.

\begin{figure}
 \epsscale{1.0}
\plotone{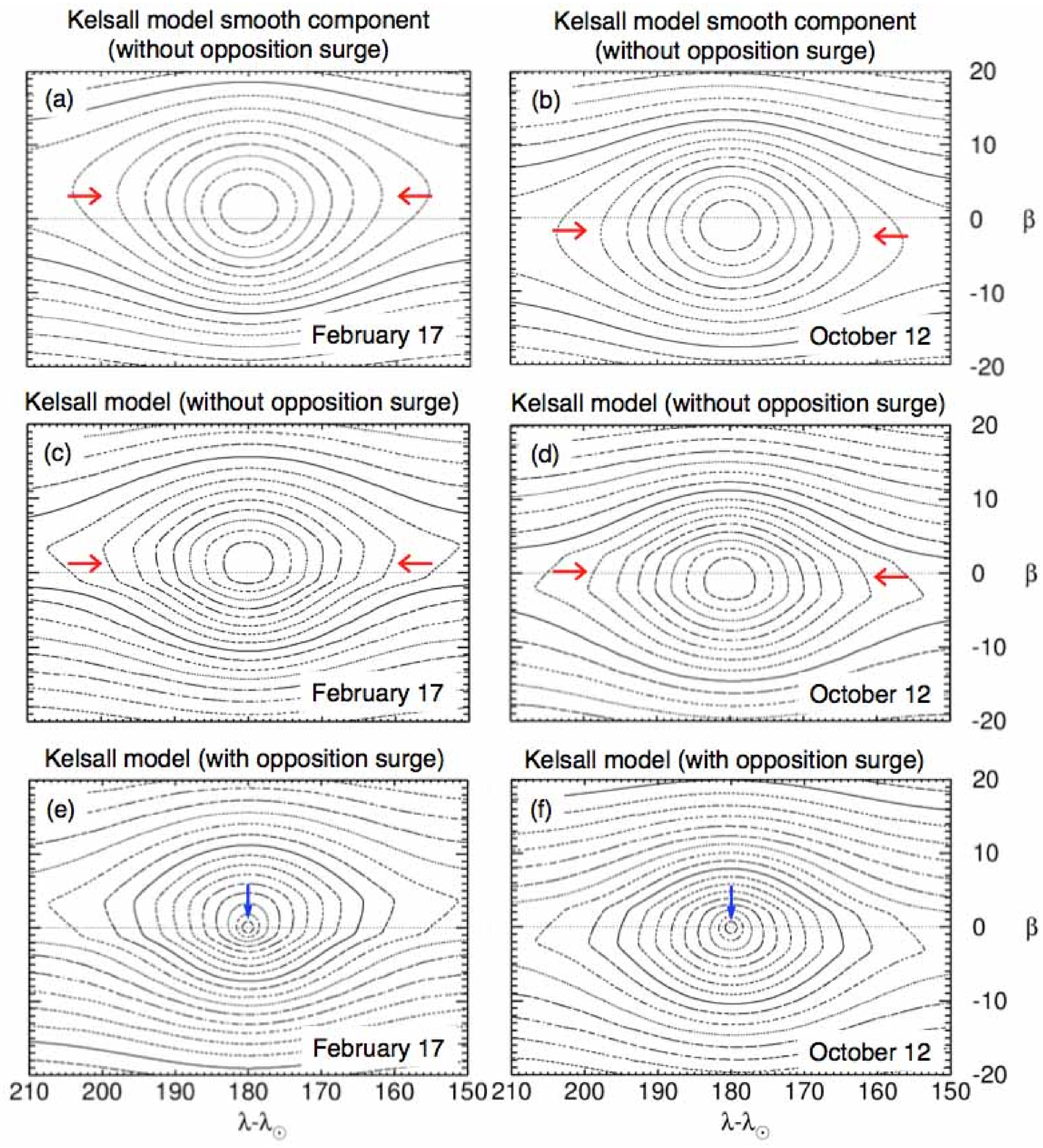}
  \caption{An isophoto map, in arbitrary units, of the Gegenschein at days from
 two different seasons  based on the infrared 3-D model
 \citep{kelsall1998}. We applied only the smooth component in the top
 panels and all components including dust bands in the middle and bottom
 panels. The original scattering phase function of
 \citet{hong1985} was applied in the top and middle panels, while the
 modified  scattering phase function was applied in the bottom two
 panels. }\label{fig:kelsall}
\end{figure}

Next, we applied the infrared model of
\citet{kelsall1998}. Because the model was constructed to fit the data
in near- and far-infrared wavelengths, no scattering phase function was
provided for optical wavelengths. We again applied the scattering phase
function in \citet{hong1985}. The upper panels in Figure
\ref{fig:kelsall} show the results using only the smooth component of the
infrared model, while the middle panels show the results using all
the components. From the comparison between top and middle panels,
it is clear that the angular shape is associated with fine-scale
structure, in particular, the dust-band components. Both the north--south
asymmetry and the  angular edge along the east--west direction were
reproduced by the model (Figures \ref{fig:kelsall}(c) and  (d)).
It is interesting to note that the model was constructed based on
infrared observations and yet fit  the optical zodiacal light as
well. In addition, the infrared observations covered the
zodiacal emission at solar elongation of 64\arcdeg--124\arcdeg.
Both the observed wavelengths and the solar elongation angle of
the infrared observations  differ from those of our
observations. Nevertheless, the infrared model matches the Gegenschein in
the visible wavelength range.

\begin{figure}
 \epsscale{1.2}
\plottwo{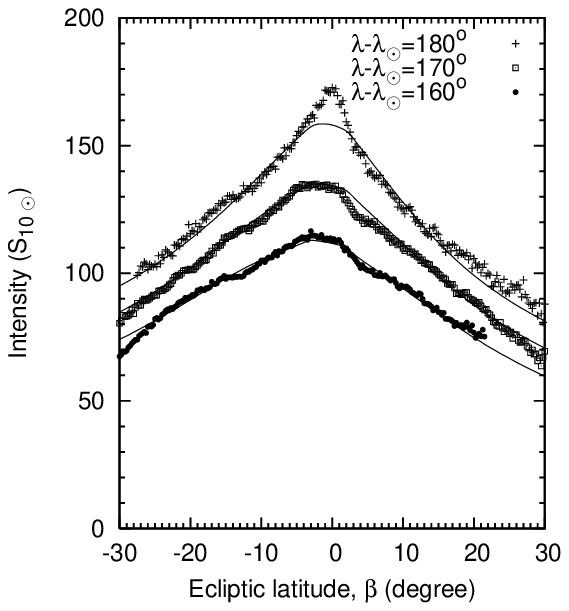}{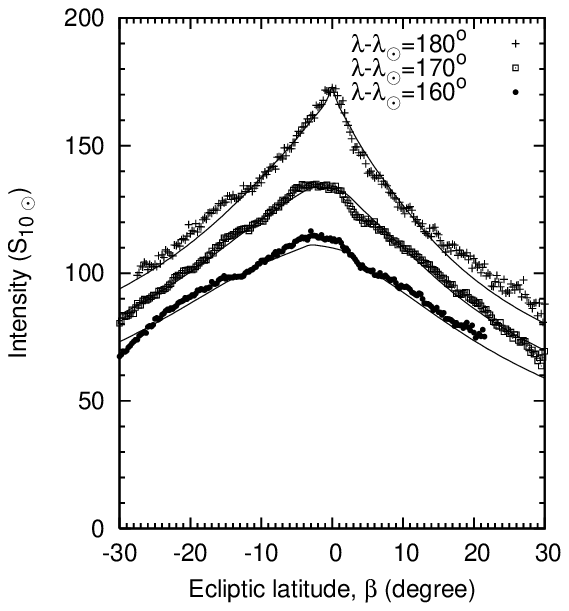}
  \caption{Comparison between the observed brightness profiles with
 those obtained from the model (solid lines). We applied the original
 scattering phase function of \citet{hong1985} assuming $\nu=1.3$  in
 the left  panel and a modified scattering phase function of
 \citet{hong1985} with 15\% enhancement at opposition in the right panel.
}\label{fig:comp_lat}
\end{figure}

In Figures \ref{fig:kelsall} and \ref{fig:contour}, we would like
to draw attention to the antisolar point, where the brightness spike
is clearly seen in our observational images but not visible in the model
image. The difference is also seen in Figure \ref{fig:comp_lat}, where
the observed brightness profile around the antisolar point does not
match  the observed one.
This is probably because the backscattering part of the scattering phase
function is inadequate for reproducing the observed spikes.
Since the scattering phase function has so far been derived using low-resolution data, it is likely that the backscattering part within
$\sim$1\arcdeg~ has not been considered yet.
For this reason, we modified the scattering phase function by multiplying
it by a term used in \citet{kaasalainen2003},
\begin{equation}
\Phi'(\theta)=\left\{1 + A~\exp\left(-\frac{\pi-\theta}{B}\right)\right\}~\Phi(\theta),
\end{equation}
\noindent where $A$ and $B$ are fitting parameters and $\Phi(\theta)$ is
the original scattering phase function in \citet{hong1985}. We derived
$A=0.15$ and $B=0.021$ (or 1.2\arcdeg) to match the observed images.
With the modified scattering phase function, we fit the surface
brightness profiles (see Figure \ref{fig:kelsall} (e)--(f) indicated by
arrows, and also Figure \ref{fig:comp_lat}).

\section{Discussion}

The question we have to address here is whether the brightest spot on the
Gegenschein was caused by  scattered light from  interplanetary
dust particles, by an instrumental effect (i.e., uneven sensitivity of CCD
pixels or vignetting), or by other light sources such as stars and
galaxies. Since we detected the brightness enhancement in different
positions on the CCD, it is unlikely that the brightness enhancement was
caused by instrumental effects.  In addition, because the antisolar
position moves relative to stars and galaxies at a rate of ~1\arcdeg/day, we can distinguish zodiacal light components from
stellar objects by comparison with  images taken on different
nights. For these reasons, it is clear that the brightness enhancement
 originates from the interplanetary dust cloud, because it always
appears at the antisolar point.

Our observational result indicates that the volume scattering phase function
sharply increases close to a phase angle of zero (that is, phase angle =
180\arcdeg - scattering angle $\theta$=0\arcdeg).
The sharp increase (hereafter referred to as the opposition spike) is a common
phenomenon inherent to airless bodies  in the Solar System, such as the
Moon, asteroids, satellites of the major planets, Saturn's rings
(Mishchenko 1993), and Kuiper Belt objects (Sheppard and Jewitt 2002);
it is characterized by a nonlinear increase in surface brightness as
the phase angle decreases to zero. Through this effect, the amplitude and the
width of the scattered light change significantly within 1\arcdeg--5\arcdeg.
There are two main physical mechanisms to explain the opposition effect
on a surface covered with  regolith:  shadow-hiding and
coherent backscattering enhancement.  Shadows directly opposite the Sun
are  hidden by the scatterers, and the absence of visible shadow
makes the area brighter on average (the shadow-hiding effect). On the other hand, coherent backscattering is an effect that describes the
appearance of an intensity cone when photons traveling in a time-reversed path self-interfere constructively in the backscattered
direction.

The opposition effect has been detected to occur on airless bodies in
the Solar System. Shadow-hiding and coherent backscattering enhancement
have been proposed as reasons for the opposition effect observed in
regolith layers \citep{muinonen2002}. It is 
interesting to note that the opposition effect was detected in the
interplanetary dust cloud from our observations. This evidence may
suggest that scattering by a single interplanetary dust particle induces
an opposition surge. Shkuratov \& Helfenstein (2001) predicted from
their model, in which the aggregate structure of regolith is considered,
that a single regolith particle could exhibit its own opposition
effect. Hapke (2002) also emphasized that multiple scatterings can occur
between different parts of a single particle. 
\citet{min2010} showed that the large ($>$ 100 \micron) dust grains
covered by regolith particles are responsible to the strong backward
scattering observed in the Fomalhaut debris disk \citep{LeBouquin2009}.
Further researches
involving multiband observations, polarimetric observation and
theoretical studies are required to study the mechanisms responsible for
the opposition effect observed in the Gegenschein. Although we cannot
conclude whether the opposition effect in the Gegenschein is caused by the
shadow-hiding effect or interference or both, we can safely conclude
that the effect is induced by scatterers whose sizes are significantly
larger than the optical wavelength.

The albedo of  interplanetary dust particles leads us to consider their
origin. This has been investigated in literature; \citet{cook1978} studied the
albedo using  micrometeoroid experiment data onboard Helios A and
Pioneer 10 as well as  zodiacal light observational data and derived
an albedo ranging from 0.07 at 0.1 AU from the Sun through 0.006 at 1 AU
and down to 0.0001 at 3.3 AU. Later, Hanner (1980) reanalyzed the data to
obtain a mean geometric albedo of 0.24 for  dust particles near 1
AU. Dumont \& Levasseur-Regourd (1988) and Renard et al. (1995) studied
 zodiacal light and  zodiacal emission by using the node of lesser
uncertainty method, and they obtained an albedo at 1 AU of 0.08--0.15 at a
scattering phase function of 90\arcdeg, which  translates into a
geometric albedo of 0.16--0.30 with the scattering phase function by Hong
(1985). Lumme \& Bowell (1985) derived an albedo of 0.04 using an
empirical polarimetric method. All these measurements are inconsistent with
each other and have not yet led to a discussion about the origin.

Our observational data can be used to derive an accurate albedo value in
combination with the infrared model. Because our observed quantity is
obtained at the opposition direction from the Sun, it gives
straightforward input for deriving the geometric albedo of
interplanetary dust particles around the Earth orbit (Appendix A.) 
because it is
defined at zero phase angle. In addition, the good consistency between the
optical and infrared
spatial distributions of dust particles suggests that the effective
size of dust grains in the optical wavelength region is almost identical to that at
infrared wavelengths.  This result supports the idea that large grains
($>$10 $\mu$m) are responsible for the scattering cross section of
zodiacal light \citep{giese1976}.

In light of this large-particle
hypothesis, let us derive the geometric albedo. We rewrite
Eq. (1) at the antisolar point using the geometric
albedo $A_p$(180\arcdeg) \citep{hanner1981} as follows:
\begin{eqnarray}
I_{ZL}(180\arcdeg, 0\arcdeg)=
\int_{0}^{\infty}
F(r_h) n(x,y,z) \frac{A_p(180\arcdeg)}{\pi} \,dl.
\end{eqnarray}Because $n(x,y,z)$ is given by the DIRBE infrared model while the intensity
of zodiacal light at  opposition is obtained by our observation
($175\pm 15$ S$_{\odot10}$), we can deduce a mean geometric albedo of
$A_p(180\arcdeg)=0.07\pm 0.01$. The DIRBE infrared model also suggests
that about 7\% of the geometric cross-sectional area accounts for asteroidal
particles, particularly the inner dust band from the Karin collisional family
(whose mean geometric albedo is 0.20--0.22; \cite{harris2009,usui2011})
and the circumsolar resonance ring possibly originating from the asteroidal
belts. Given that 7\% of the geometric cross section is accounted for
by particles with $A_p(180\arcdeg)=0.20$--0.22, we found that the remaining
zodiacal light (the smooth component in the DIRBE model) should have a
geometric albedo of $0.06 \pm 0.01$.

Our result is consistent with the fact that the micrometeorites
accumulated around Earth orbit have low albedo ($<$0.1;
\citet{brownlee1978,hanner1980}). Among Solar System minor objects,
comet nuclei and C- and D-type asteroids show low albedo, that is, 0.02--0.06
for comet nuclei \citep{lamy2004} and $0.07\pm 0.04$ for
C-type asteroids \citep{usui2011}. Recent dynamical simulations of
interplanetary dust particles suggest that 85\%--95\% of the observed
mid-infrared emission is produced by particles from Jupiter-family
comets and $<$10\% by dust from long-period comets \citep{nesvorny2010}.
\citet{jewitt2013} conclude that  $<$3\% of the interplanetary dust
particles are produced by events like P/2010 A2 probably caused by
breakup of an asteroid \citep{snodgrass2010,jewitt2011}.
On the contrary, \citet{tsumura2010} insist that high-albedo 
particles  dominate near Earth orbit based on their spectroscopic
observation at near-infrared wavelengths.
Although further investigation is required on the fraction of asteroidal
particles with respect to cometary particles, it seems plausible that
majority of interplanetary dust particles have a composition similar to
carbonaceous chondrites rather than siliceous particles.


\section{Summary}

In this paper, we outlined the instrumentation, the data reduction, and the
observations of the Gegenschein using the hand-made observation system
WIZARD. Our main findings are as follows:

\begin{itemize}
\item A narrow brightness enhancement was superimposed on the smooth
gradient of the Gegenschein at the antisolar point. Hence, the
position of  maximum brightness exists at the antisolar point throughout
the year.

\item The Gegenschein morphology changed according to the orbital motion
of the Earth. It is well produced by a model constructed using infrared
observations by COBE/DIRBE.
\end{itemize}

We hypothesized that the zero phase angle spike is caused by coherent
backscattering and/or shadow-hiding effects on the rough surfaces of
individual dust particles. The above results indicate
that the effective sizes of interplanetary dust particles responsible
for optical zodiacal light and infrared zodiacal emission are much larger
than those corresponding to the observed wavelengths.

Finally, we derived the geometric albedo of interplanetary dust
particles using our observational result and found that it is $0.06\pm 0.01$
for the smooth component of the zodiacal cloud. The derived albedo is in
accordance with that of the dark micrometeorites accumulated around earth orbit and the observed cometary
dust particles. We concluded that chondritic particles are dominant
near Earth space.

\acknowledgments
We would like to express out greatest gratitude to the people who have
helped and supported us throughout our project. In particular, we are
grateful to Prof. Seung Soo Hong for his continuous encouragement and
advices from conceptual inception of the project to these days. We 
also wish to thank stuffs at IRTF and Subaru (NAOJ), especially, Alan
Tokunaga, George Koenig, Kazu Sekiguchi, Masao Nakagiri, and Akihiko
Miyashita, for their tremendous contribution and support towards the
long-term observations at Mauna Kea. The laboratory experiments were
made at University of Tokyo.
The control software of COGITO was
developed by Takehiko Wada (ISAS), Yosuke Ohno (RIKEN) and Hitoshi
Muller (Musashino Art University), and modified by Ryosuke Nakamura
(NICT). The optical system was designed and fabricated by Genesia
Corporation, Japan. Shoko Ohtsuki (Senshu University), Yuki Sarugaku
(ISAS), Naoya Miura, Hideo Sagawa, Seitaro Urakawa  helped us. The
performance test at a laboratory was supported by a grant of Seoul
National University. MI and HY are supported by National Research
Foundation of Korea (NRF) grant funded by the Korea Government (MEST)
(No. 2012R1A4A1028713).

\appendix
\section{Appendix A: Line of Sight Depth of Zodiacal Light and Zodiacal
Emission}
\begin{figure}
 \epsscale{0.75}
\plotone{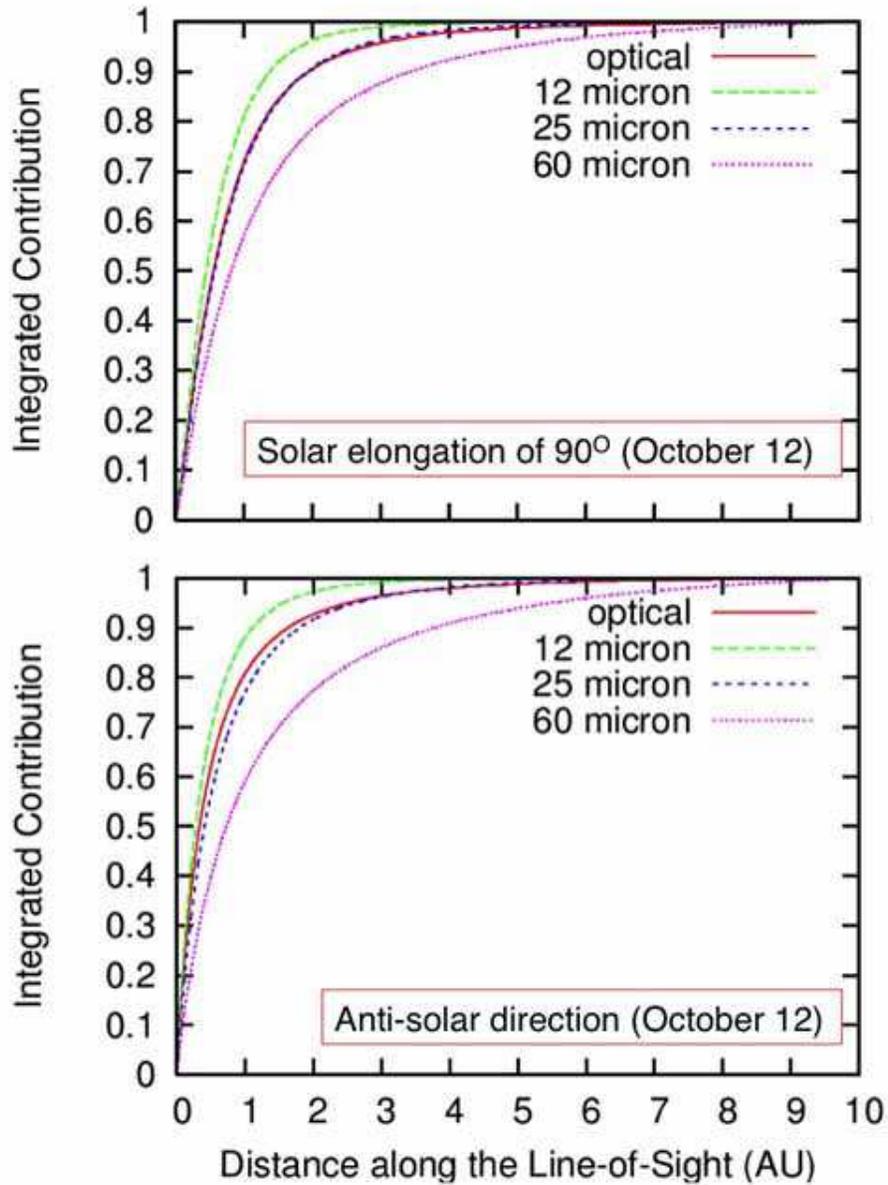}
  \caption{The line of sight depth of optical zodiacal light and
 infrared zodiacal emission at at $\lambda-\lambda_{\odot}=90$\arcdeg~ and
$\lambda-\lambda_{\odot}=180$\arcdeg~ on the ecliptic plane.
}\label{fig:append}
\end{figure}

Because the intensities of optical zodiacal light and infrared zodiacal
emission  are quantities integrated along the line of sight, it is
difficult to specify the particles we observe. We consider the
contribution of the elements at different distance from the observers to
the total flux of optical zodiacal light and infrared zodiacal
emission. Figure \ref{fig:append} show the contribution in different
wavelengths at $\lambda-\lambda_{\odot}=90$\arcdeg~ and
$\lambda-\lambda_{\odot}=180$\arcdeg~ on the ecliptic plane.
We found that the line of sight depth becomes shallowest at 12\micron~
because hot dust particles ($\sim$280 K) near Earth orbit are efficient
infrared emitters. On the other hand, distant particles contribute well
at longer wavelength. The depth in the optical wavelength is similar to
that at 25\micron. It seems that we derived the albedo  of
interplanetary dust particle within $\sim2-3$ AU from the 
Earth.

\clearpage

\clearpage

\clearpage

\clearpage

\clearpage

\clearpage

\clearpage

\clearpage

\clearpage

\clearpage

\clearpage

\clearpage

\clearpage

\clearpage


\clearpage

\clearpage


\begin{table}
\begin{center}
  \caption{Summary of the WIZARD performance}
    \begin{tabular}{|l|r|}
     \hline
     Field-of-view & 49$\arcdeg$$\times$98$\arcdeg$\\
     \hline
     Pixel resolution & 1.435$\arcmin$\\
     \hline
     Gain factor & 0.825 (e$^-$/ADU)\\
     \hline
     Readout noise & 20--25 (e$^-$)\\
     \hline
     Stability of the CCD zero level & $\lesssim$10 (e$^-$) or\\
     & $\lesssim$0.6\% of the Gegenschein brightness\\
     \hline
     Vignetting & 30 \% (at 47$\arcdeg$ from the optical center)\\
     \hline
     Pixel-to-pixel responsivity variation & $\lesssim$0.3\% \\
     \hline
    \end{tabular}
\end{center}
\end{table}

\clearpage


\begin{deluxetable}{rlrrrrrr}

\tablecaption{Observation summary}
\tablewidth{0pt}
\tablehead{ \colhead{} &
\colhead{Date} & \colhead{UT$^a$} & \colhead{$r_h{^b}$} &
\colhead{$\lambda_{\oplus}{^c}$} & \colhead{$Z^d$} & \colhead{$l^e$} &
\colhead{$b^f$}}

\startdata

2003 & March 01 & 10:05 & 0.991 & 160.36 & 14.04 & 241 & +55\\
     & & 10:29 & 0.991 & 160.37 & 12.21 & 241 & +55\\
     & March 02 & 09:35 & 0.991 & 161.34 & 18.99 & 242 & +55\\
     & & 09:58 & 0.991 & 161.36 & 15.25 & 242 & +55\\
     & & 10:45 & 0.991 & 161.39 & 12.82 & 242 & +55\\
     & & 11:35 & 0.991 & 161.42 & 19.41 & 242 & +55\\
     & August 25 & 09:58 & 1.011 & 331.75 & 31.33 & 49 & $-$50\\
     & & 10:26 & 1.011 & 331.77 & 30.68 & 49 & $-$50\\
     & & 10:49 & 1.011 & 331.78 & 31.28 & 49 & $-$50\\
     & August 27 & 10:06 & 1.010 & 333.69 & 30.28 & 52 & $-$51\\
     & & 10:29 & 1.010 & 333.70 & 30.01 & 52 & $-$51\\
2004 & February 15 & 09:41 & 0.988 & 145.98 & 14.87 & 223 & +46\\
     & & 10:00 & 0.988 & 145.99 & 11.06 & 223 & +46\\
     & February 16 & 08:23 & 0.988 & 146.94 & 32.64 & 224 & +47\\
     & & 09:01 & 0.988 & 146.96 & 23.86 & 224 & +47\\
     & & 09:41 & 0.988 & 146.99 & 15.03 & 224 & +47\\
     & & 10:05 & 0.988 & 147.01 & 10.39 & 224 & +47\\
     & February 17 & 09:09 & 0.988 & 147.98 & 22.17 & 225 & +47\\
     & & 09:27 & 0.988 & 147.99 & 18.18 & 225 & +47\\
     & & 09:44 & 0.988 & 148.00 & 14.58 & 225 & +47\\
     & April 21 & 09:28 & 1.005 & 211.55 & 34.34 & 327 & +48\\
     & & 10:05 & 1.005 & 211.58 & 32.07 & 327 & +48\\
     & & 10:24 & 1.005 & 211.59 & 31.87 & 327 & +48\\
     & & 10:42 & 1.005 & 211.60 & 32.30 & 327 & +48\\
     & April 22 & 09:24 & 1.005 & 212.52 & 35.01 & 328 & +47\\
     & & 09:42 & 1.005 & 212.54 & 33.52 & 328 & +47\\
     & September 12 & 11:53 & 1.006 & 349.98 & 33.37 & 77 & $-$59\\
     & & 12:12 & 1.006 & 350.00 & 36.79 & 77 & $-$59\\
     & September 13 & 11:25 & 1.006 & 350.94 & 28.71 & 79 & $-$59\\
     & & 11:45 & 1.006 & 350.95 & 31.80 & 79 & $-$59\\
     & & 12:04 & 1.006 & 350.96 & 35.12 & 79 & $-$59\\
     & October 13 & 08:52 & 0.998 & 20.29 & 21.87 & 133 & $-$54\\
     & & 09:13 & 0.998 & 20.31 & 17.82 & 133 & $-$54\\
     & October 14 & 09:22 & 0.997 & 21.30 & 15.95 & 134 & $-$54\\
     & & 09:41 & 0.997 & 21.32 & 13.19 & 134 & $-$54\\
     & & 09:57 & 0.997 & 21.33 & 11.79 & 134 & $-$54\\
     & & 10:13 & 0.997 & 21.34 & 11.58 & 135 & $-$54\\
     & November 11 & 09:59 & 0.990 & 49.29 & 2.77 & 163 & $-$34\\
     & & 10:18 & 0.990 & 49.30 & 3.70 & 163 & $-$34\\
     & & 10:36 & 0.990 & 49.31 & 7.53 & 163 & $-$34\\
     & November 12 & 10:16 & 0.990 & 50.30 & 3.14 & 164 & $-$34\\
     & & 10:33 & 0.990 & 50.32 & 6.74 & 164 & $-$34\\
     & & 10:49 & 0.990 & 50.33 & 10.41 & 164 & $-$34\\
     & November 13 & 10:19 & 0.989 & 51.31 & 3.54 & 165 & $-$33\\
     & & 10:52 & 0.989 & 51.34 & 11.02 & 165 & $-$33\\
2006 & November 18 & 08:47 & 0.988 & 55.77 & 18.76 & 168 & $-$29\\
     & & 09:05 & 0.988 & 55.78 & 14.53 & 168 & $-$29\\
     & November 23 & 08:41 & 0.987 & 60.81 & 20.38 & 171 & $-$25\\
     & & 08:59 & 0.987 & 60.83 & 16.16 & 171 & $-$25\\
     & November 27 & 12:01 & 0.987 & 65.00 & 26.21 & 173 & $-$22\\
\enddata
\tablecomments{
$^a$Median time of exposure.\\
$^b$Heliocentric distance of the Earth in {AU}.\\
$^c$Ecliptic longitude of the antisolar point with the equinox of {J2000} in degrees.\\
$^d$Zenith distance of the antisolar point in degrees.\\
$^e$Galactic longitude of the antisolar point in degrees.\\
$^f$Galactic latitude of the antisolar point in degrees.
}
\label{table:obs}
\end{deluxetable}

\end{document}